\documentclass[10pt,twocolumn,pra,aps,superscriptaddress, floatfix]{revtex4-2}
\usepackage[english]{babel}
\usepackage{amsmath}
\usepackage{graphicx}
\usepackage{soul}
\usepackage[colorlinks=true, allcolors=blue]{hyperref}
\usepackage{mathrsfs}

\usepackage{csquotes}
\usepackage{xcolor}
\usepackage{soul}
\usepackage{orcidlink}
\usepackage{ragged2e}
\date{\today}

\begin{document}
\title{Driven Magnon-Photon System as a Tunable Quantum Heat Rectifier }
\author{C. O. Edet\orcidlink{0000-0001-7762-731X}}
\affiliation{Faculty of Electronic  Engineering  \& Technology, Universiti Malaysia Perlis, 02600 Arau, Perlis, Malaysia }
\email{collinsokonedet@gmail.com}
\author{K. S\l{}owik \orcidlink{0000-0003-1314-7004}}
\affiliation{Institute of Physics, Faculty of Physics, Astronomy and Informatics, Nicolaus Copernicus University in Toruń, Grudziadzka 5/7, 87-100 Toruń, Poland}
\author{N. Ali \orcidlink{0000-0002-9348-0714}}
\affiliation{Faculty of Electronic  Engineering  \& Technology, Universiti Malaysia Perlis, 02600 Arau, Perlis, Malaysia }
\author{M. Asjad \orcidlink{0000-0001-6895-3332}}
\affiliation{Department of Physics, Simon Fraser University, Burnaby, B.C., Canada V5A 1S6}
\author{O. Abah \orcidlink{0000-0003-0193-4860}}
\affiliation{School of Mathematics, Statistics and Physics, Newcastle University, Newcastle upon Tyne NE1 7RU, United Kingdom}.

\begin{abstract}
Controlling heat flow at the quantum level is a key challenge for next-generation quantum technologies, including thermal management and quantum information processing. Here, we investigate quantum heat transport in an asymmetrically driven hybrid magnon-photon system in contact with two thermal baths at different temperatures. 
We demonstrate that external driving of the magnonic subsystem provides a versatile control knob for tailoring steady-state heat currents and their asymmetry.
We identify the mechanisms governing thermal rectification in the hybrid system: we find that strong rectification emerges in the regime of weak magnon–photon hybridization combined with intense magnon driving. 
In this regime, the external drive enables control over both the magnitude and direction of the heat current, allowing the rectification parameter to be tuned across its entire physically accessible range.
\end{abstract}
\maketitle

\section{Introduction}
In the quest to miniaturize electronic devices, regulation of heat transport at the nanoscale remains a tremendous impediment to the advancement of quantum and classical computing architecture \cite{senior2020heat}. Rapid progress in quantum technology has led to the need for alternative information carriers, such as magnetic (spin) \cite{Ifmmode2004RMP,wolf2001spintronics,Poulsen2021PRL} and thermal (phonon) \cite{Saaskilahti2013PRE,Li2012RMP,roberts2011review} currents, which are being explored alongside traditional charge carriers. The thermal current, central to thermodynamics, behaves differently at the quantum scale, governed by quantum mechanics.
Moreover, minimizing heat dissipation and exploiting the control of heat currents in quantum systems are important for applications in the area of quantum information processing \cite{PtaszyifmmodePRL2019}, designing quantum thermal devices \cite{bouton2021quantum,Levy2012PRE,myers2022quantum,Gelbwaser2013PRE}, heat transistors \cite{wang2018heat, du2019quantum,guo2019multifunctional,Majland2020,Mandarino2021PRApplied,mandarino2021thermal,malavazi2024detuning,Yang2024},  and thermal rectifiers \cite{Segal2005PRL,Yan2009PhysRevB,Werlang2014,Marcos2018,goury2019reversible,kargi2019quantum,upadhyay2021heat,senior2020heat,ivander2022quantum,poulsen2022,khandelwal2023characterizing,Rajapaksha2024,Liu2024}.

Theoretical \cite{riera2019dynamically,Alexander2020,Iorio2021,Simon2021,Stevenson2021,Palafox2022,tesser2022heat,Liu2023} and experimental \cite{Seif2018,senior2020heat} studies of quantum thermal rectifiers are areas of current research.
The thermal rectifier is a device in which the direction of the heat current is reversed when there is a thermal bias due to asymmetry and non-linearity in physical systems \cite{khandelwal2023characterizing, bhandari2021thermal}. We remark, for two bosonic thermal reservoirs, a simple two-level system can act as a heat rectifier, while that is not the case for linearly coupled harmonic oscillators \cite{wu2009sufficient,Kalantar2021}.
The quantum segmented $XXZ$ chains show strong  rectification of the heat current, emanating from the asymmetry created by different on-site magnetic fields, driving field amplitude, and the coupling strength of the system with their individual baths \cite{Landi2014PRE, Ordonez-Miranda2017PRE, Balachandran2019PRE}. 
A giant rectification has recently been proposed in a one-dimensional lattice of spinless fermions based on the asymmetric interplay between strong particle interactions and a tilted potential \cite{Mendoza2024}. Heat rectifiers have been investigated on different platforms; solid state quantum circuits \cite{Balachandran2018}, coupled two-level atoms \cite{Iorio2021,Liu2024}, quantum dots \cite{Jiang2015,Aligia2020}, or superconducting circuits \cite{Xu2021}. 

Furthermore, hybrid magnonic systems have shown great potential for the development of novel quantum technologies \cite{Huebl2013PRL,Tabuchi2014PRL, lachance2019hybrid,Li2021,Potts2021}. Magnons, the collective spin excitation in yttrium iron garnet (YIG), have been coupled with microwave photons through the magnetic dipole \cite{Zhang2014PRL}, optical photons via the magneto-optical effect \cite{Osada2016PRL}, superconducting quantum circuits indirectly through the mediation of microwave cavity photons \cite{tabuchi2016quantum}, among others. A plethora of interaction mechanisms allows the engineering of coherent interactions between different systems \cite{Yuan2022}. In addition, hybrid quantum-magnonic systems provide opportunities for the development of quantum technologies useful in quantum sensing \cite{Ebrahimi2021}, quantum information processing \cite{lachance2019hybrid}, and understanding the interplay between irreversibility and quantum information in a mesoscopic quantum system \cite{Edet2024}. Motivated by the versatile nature of driven hybrid magnonic systems, we will investigate their quantum thermal applications. 

\begin{figure}[!ht]
\centering
\includegraphics[width=0.98\columnwidth]{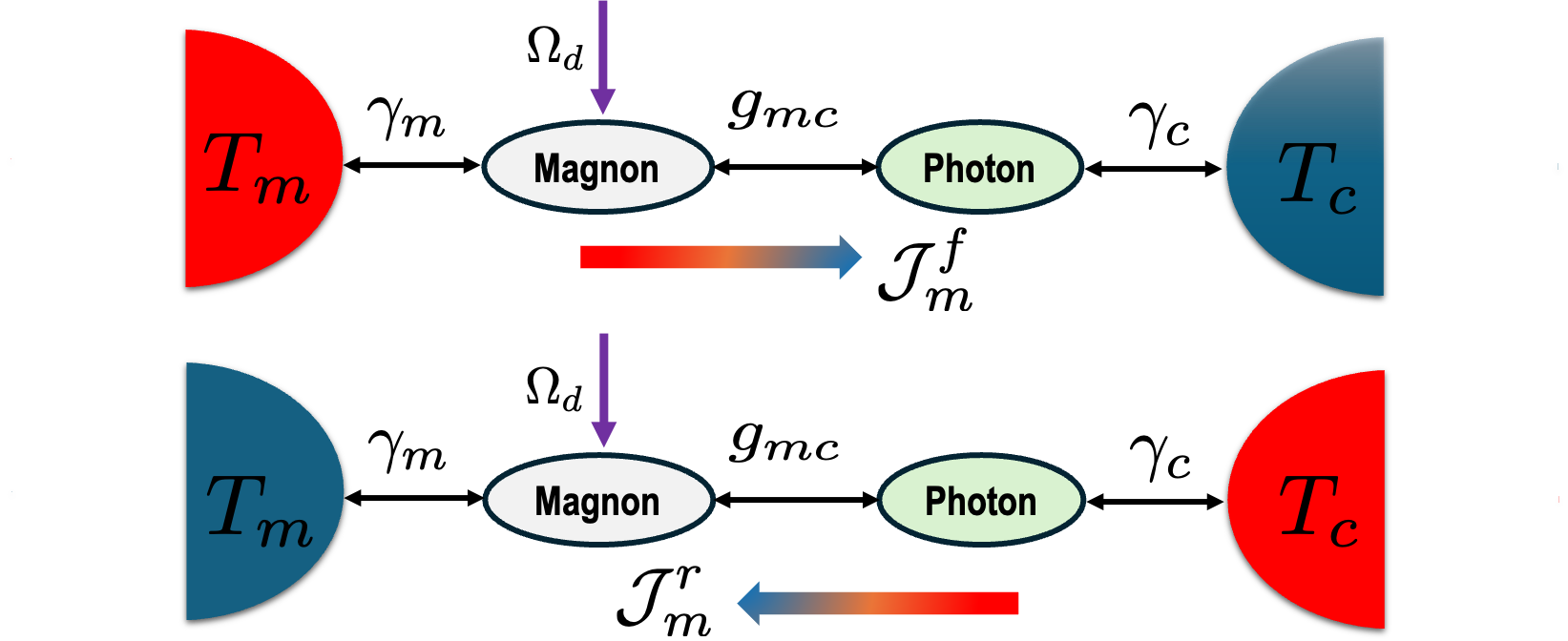} 
\caption{\justifying Heat rectification architecture of a driven hybrid magnon-photon system coupled to two different heat reservoirs at fixed temperatures ($T_m$ and $T_c$)  with system-bath couplings $\gamma_m$ and $\gamma_c$. The magnon mode is driven with a driving field of strength $\Omega_d$ and the magnon-photon coupling strength $g_{mc}$. (Top) forward-bias configuration; (bottom) reversed-bias configuration.}      
\label{figmp}
\end{figure}

In the present study, we introduce a class of tunable quantum rectifiers that utilizes the quintessential parameter of a driven magnon-photon system. Specifically, we investigate the heat current and rectification in a magnon-photon system where the magnon mode is driven; see Fig.\ref{figmp}. 
Without driving, the forward heat current $\mathcal{J}_{m}^f $ is induced by a positive temperature bias $(T_m>T_c)$, while the reverse heat current $\mathcal{J}_{m}^r$ arises from a negative temperature bias $(T_c>T_m)$. When the left-right (magnon-photon) symmetry is broken, the magnitude of the heat current $\mathcal{J}_{m}^f$ may be different with respect to the magnitude of $\mathcal{J}_{m}^r$. 
We show that asymmetry can be induced in a magnon-photon system independently of the temperature bias, if the magnonic subsystem is driven externally. The nontrivial asymmetry in the heat flow due to mode-selective interactions has not been discussed to the best of our knowledge.
We find that the presence of the magnon drive results in finite heat current even for small coupling strength. The control knob, magnon drive field enables full range of rectification in the setup.
Our analytical results for both heat currents and rectification reveal the underlying mechanism that leads to thermal rectification.

The structure of the paper is as follows. We introduce the model of the hybrid system and their dissipation as well as derive the open system dynamics evolution equations in Sec. \ref{sectm-photon}. In Sec. \ref{sectionThermo}, we show the dependence of the heat current on the parameters of the magnon-photon system, and demonstrate that heat rectification can be achieved in such a system by driving the magnon mode. In Sec. \ref{expt}, we present the experimental feasibility of the proposed quantum thermal rectifier in current state-of-the-art cavity magnonic systems. Finally,  the conclusion is given in Sec. \ref{conclusion}.

\section{Model and dissipative dynamics} \label{sectm-photon}
We consider a hybrid quantum system consisting of coherently coupled magnons and microwave photons, which are connected to two different thermal reservoirs, as shown in Fig. (\ref{figmp}). The magnons are directly driven by an external magnetic field, and couple to the cavity photons via the magnetic dipole interaction \cite{lachance2019hybrid}. 
The total Hamiltonian of the driven hybrid  magnon-photon system is given in the rotating wave approximation \cite{crescini2020magnon, Sun2021PRL, Zhang2014PRL};
\begin{eqnarray}
\hat{H}/\hbar &=& \omega_{c} \hat{c}^{\dagger} \hat{c} \,+\, \omega_{m} \hat{m}^{\dagger} \hat{m} \,+\,  g_{mc} \left(\hat{c} \hat{m}^{\dagger} + \hat{c}^{\dagger} \hat{m}\right) \nonumber\\ &+& i \Omega_d \left(\hat{m}^{\dagger} e^{-i\omega_d t}  - \hat{m} e^{i\omega_d t}\right),
\label{eq1}
\end{eqnarray}
where  $\hat{c} (\hat{c}^{\dagger})$ and  $\hat{m} (\hat{m}^{\dagger})$  are the annihilation (creation) operators for the photon and magnon modes respectively. The first two terms are the bare Hamiltonians of the microwave cavity and the magnon with frequencies $\omega_c$ and $\omega_m$, respectively. The third term denotes the interaction between the photon and the magnon with the coupling strength $g_{mc}$,  which can be adjusted by varying the direction of the bias field or the position of the YIG sphere inside the cavity \cite{Zhang2014PRL, lachance2019hybrid}. The last term represents the magnon mode drive with the amplitude strength $\Omega_d$ and the driving frequency $\omega_d$.
Making a unitary transformation of the form $\hat{U}\!=\!\text{{exp}} (i\omega_d ( \hat{m}^{\dagger} \hat{m} + \hat{c}^{\dagger} \hat{c})t)$, in a rotating frame at the magnon driving frequency $\omega_d$, the effective Hamiltonian of Eq.~(\ref{eq1}) reads
$\hat{H}_{\text{eff}}\!=\!\hat{U} \hat{H} \hat{U}^{\dagger} - i \hbar \hat{U} (\partial\hat{U}^{\dagger}/\partial t)$, 
\begin{equation}
 \hat{H}_{\text{eff}}/\hbar = \Delta_{c} \hat{c}^{\dagger} \hat{c}+ \Delta_{m} \hat{m}^{\dagger} \hat{m} +  g_{mc} (\hat{c} \hat{m}^{\dagger}+ \hat{c}^{\dagger} \hat{m})    + i \Omega_d (\hat{m}^{\dagger}  - \hat{m} )
\end{equation}
where $\Delta_{c}\!=\!\omega_c - \omega_d$ and  $ \Delta_{m}\!=\!\omega_m - \omega_d$ represent the detuning of the photonic and magnonic modes from the drive, respectively. 




Next, we move to present and discuss the dissipative dynamics of the system. The heat current is evaluated in the stationary state within the 
 Lindblad-Gorini-Kossakowski-Sudarshan (GSKL)  Makovian master equation framework \cite{gorini1976completely, lindblad1976generators}. We assume Markovian bosonic reservoirs and weak coupling between the reservoirs and devices. Hence, the evolution equation for the hybrid magnon-photon system is presented with the adjoint master equation for the hybrid magnon-photon setup  \cite{breuer2002theory}
\begin{equation}\label{adjointmaster}
d \hat{O}_{H}/dt = \frac{i}{\hbar} [\hat{H}_{\text{eff}}, \hat{O}_{H}(t)] + \mathcal{L}(\hat{O}_{H}) 
\end{equation}
where  $\hat{O}_{H}(t)$ is an arbitrary Heisenberg-picture operator at time $t$, and the second term,
\begin{equation}
\mathcal{L}(\hat{O}_{H}) = \sum_{k} \gamma_{k} (n_k+1) \mathcal{D}[\hat{O}_k] (\hat{O}_{H}) + \sum_{k} \gamma_{k} n_k \mathcal{D}[\hat{O}_k^\dagger](\hat{O}_{H}) 
\label{Lindbladian}
\end{equation} 
where $\mathcal{D}[\hat{O}_k](\hat{O}_{H}) = \hat{O}_k^{\dagger} \hat{O}_{H} \hat{O}_k -\frac{1}{2}   \hat{O}_{H}\hat{O}_k^{\dagger} \hat{O}_k-\frac{1}{2}   \hat{O}_k^{\dagger} \hat{O}_k\hat{O}_{H}$ 
is the adjoint dissipator \cite{landi2018quantum}, which is a superoperator acting on observables $\hat{O}_{H}$ and $\hat{O}_k\in \{\hat{m}, \hat{c}\}$ are the jump operators which describe how the magnon-photon system exchanges energy or information with it environment. Here,  $\gamma_{k}$ are the strength of the interaction with the reservoirs, 
$n_{k} (\omega_k, T_k)\!=\!(\exp(\hbar \omega_k/k_B T_k) - 1)^{-1}$, $k\in\{m, c\}$ are the mean number of excitations in the reservoir damping the mode, $T_k$ is the temperature of the coupled bath, and $k_B$ is the Boltzmann constant. 
The dissipator term represents the interaction of the system with the two environmental heat baths and accounts for the exchange of photons/magnons with the coupled heat reservoirs. For magnons,the primary heat bath is the phonon environment, with dissipation governed by Gilbert damping (spin-lattice relaxation). 
Photons, on the other hand, interact with an external thermal environment through the microwave resonator's physical connections with a photonic environment.
The two thermal reservoirs can be tuned independently while the cross talk can be suppressed by careful frequency matching \cite{Zhang2014PRL}.

A study of the dynamics of the system can provide analytical expressions for the heat currents and the rectification parameter, both expressed, via expectation values of the magnonic and photonic excitation numbers. To derive these expressions, we start with the adjoint master equations \eqref{adjointmaster}, to evaluate the averages
of the time derivatives of the relevant operators.  Solving the resulting equations of motion for $\hat{O}_H\in \{\hat{m},\hat{c},\hat{c}^\dagger \hat{c}, \hat{m}^\dagger\hat{m},\hat{c}^\dagger\hat{m}\}$, we obtain the averages of the operator expectation value derivatives
\begin{align}\label{firstmoment}
    \frac{d\langle \hat{c} \rangle}{dt} &= -i \Delta_c \langle \hat{c} \rangle -ig_{mc} \langle m\rangle - \frac{\gamma_c}{2} \langle \hat{c} \rangle,\\
    \frac{d\langle \hat{m} \rangle}{dt} & =  - i\Delta_m \langle \hat{m} \rangle - i g_{mc} \langle \hat{c} \rangle -\frac{\gamma_m}{2} \langle \hat{m} \rangle + \Omega_d.
\end{align}
The quantum heat transport in the hybrid setup is governed by the second moments of the mode operators, with their derivatives presented as follows:
\begin{align} \label{secondmoment}
    \frac{d \langle \hat{c}^{\dagger} \hat{c}\rangle}{dt} &= i g_{mc} (\langle \hat{m}^{\dagger} \hat{c}\rangle - \langle \hat{c}^{\dagger} \hat{m}\rangle) +\gamma_c (n_c- \langle \hat{c}^{\dagger} \hat{c}\rangle), \\
    \frac{d \langle \hat{m}^{\dagger} \hat{m}\rangle}{dt} & =   i g_{mc} (\langle \hat{c}^{\dagger} \hat{m}\rangle - \langle \hat{m}^{\dagger} \hat{c}\rangle)   + \Omega_d (\langle \hat{m}^{\dagger} \rangle \nonumber + \langle \hat{m}\rangle) \\&+ \gamma_m (n_m -\langle \hat{m}^{\dagger} \hat{m}\rangle),
\end{align}
This demonstrates that the population of each mode is explicitly impacted by $\langle \hat{m}^{\dagger} \hat{c}\rangle$, representing the excitation hopping between the magnon and photon modes. The expectation value of the hopping excitation is given by
\begin{align} \label{correlation}
\frac{ d \langle \hat{c}^{\dagger} \hat{m} \rangle}{dt} &= i (\Delta_c -\Delta_m)  \langle \hat{c}^{\dagger} \hat{m} \rangle + i g_{mc} ( \langle \hat{m}^{\dagger} \hat{m} \rangle -  \langle \hat{c}^{\dagger} \hat{c} \rangle) \nonumber \\&+ \Omega  \langle \hat{c}^{\dagger} \rangle - \left(\frac{ \gamma_{c}}{2} +\frac{\gamma_{m}}{2}\right)  \langle \hat{c}^{\dagger} \hat{m} \rangle.
\end{align}
Solving the complete set of Eqs. (\ref{firstmoment}) - (\ref{correlation}) for the steady states by setting $d \langle...\rangle/dt\!=\!0$, symmetric bath-system coupling ($\gamma_m \!=\!\gamma_c\!=\!\gamma$) and detunings ($\Delta_m\!=\!\Delta_c\!=\!\Delta$), we obtain the stationary expectation values as;
\begin{equation} \label{c2moment}
\langle \hat{c}^{\dagger} \hat{c} \rangle =\frac{2 g_\text{mc}^2 (n_\text{m}-n_\text{c})}{\gamma ^2+4 g_\text{mc}^2}+\frac{ g^2_\text{mc} \Omega_{\rm d}^2 }{\Xi }  + n_{\rm c} 
\end{equation}
 \begin{equation}\label{m2moment}
\langle \hat{m}^{\dagger} \hat{m} \rangle=\frac{\Omega_{d}^{2}\left(\Delta^{2}+\frac{\gamma^{2}}{4}\right)}{ \Xi} +\frac{2 g_\text{mc}^2 (n_\text{c}-n_\text{m})}{\gamma ^2+4 g_\text{mc}^2}+ n_{\rm m },
\end{equation}
\begin{equation}\label{cmmoment}
    \langle \hat{c}^{\dagger} \hat{m} \rangle= i\left(g_\mathrm{mc} \Omega_{d}^{2}\frac{(\frac{\gamma}{2}+i\Delta)}{\Xi}+\frac{g_\mathrm{mc} \gamma}{\gamma^{2}+4g_\mathrm{mc}^{2}}\left(n_{m}-n_{c}\right)\right),
\end{equation}
where $\Xi=\left(g^{2}_{mc}-\Delta^{2}+\frac{\gamma^{2}}{4}\right)^{2}+\Delta^{2}\gamma^{2}$.
\section{Thermodynamic analysis}\label{sectionThermo}
Here, we analyze the heat transport and rectification coefficient of the driven-dissipative magnon-photon system. Specifically, we will focus on how these quantities are affected by the system parameters.
\subsection{Heat Current} \label{SSheatcurrent}
The steady-state heat current flows through the device when n asymmetry is imposed on the setup. This is typically achieved with a temperature bias. As we show below, asymmetry and the resulting heat current can be alternatively realized with a magnonic drive.

\begin{figure*}[!]
\includegraphics[width= 1.9\columnwidth]{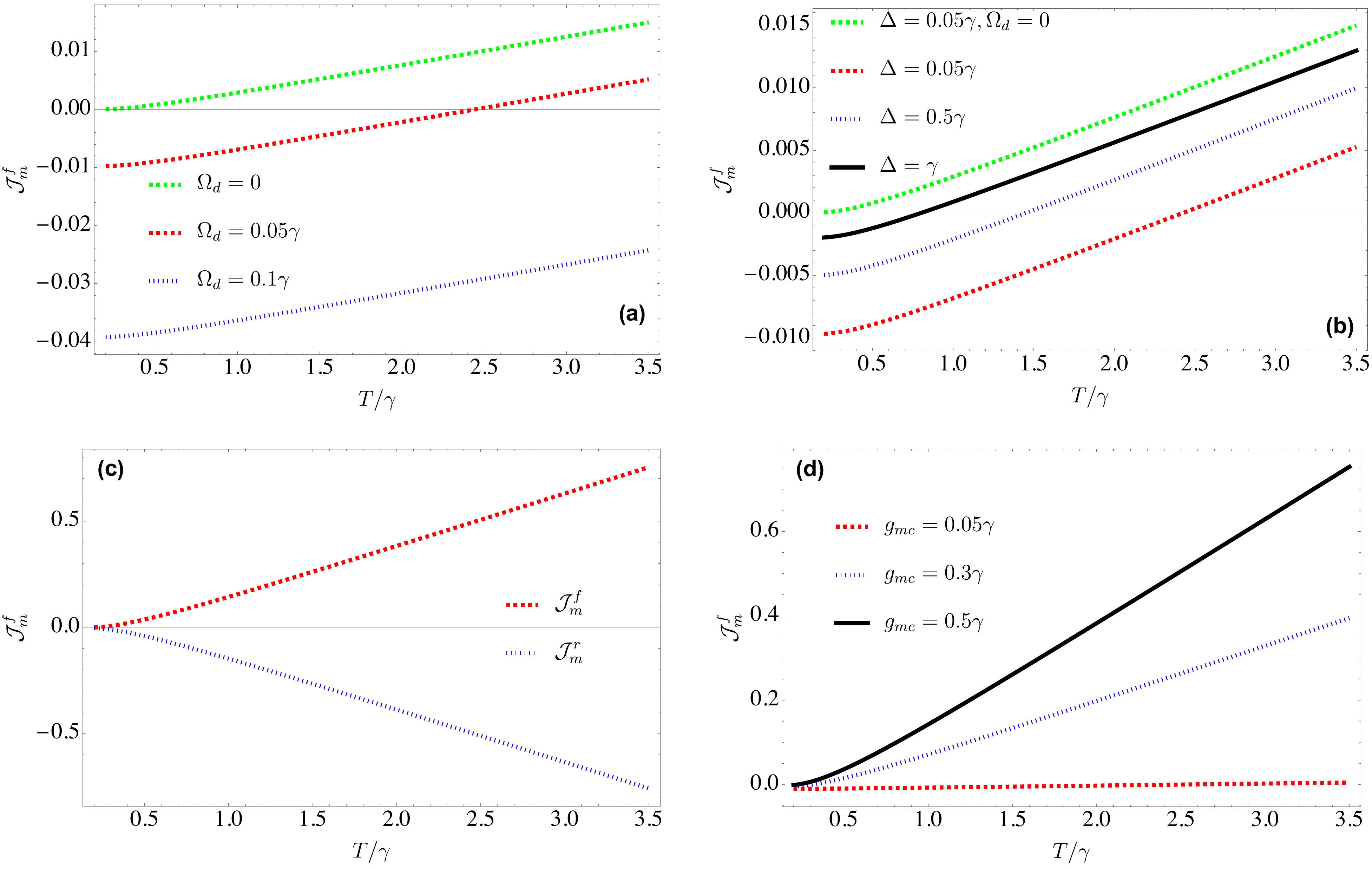} 
    \caption{\justifying Steady-state magnon heat current $\mathcal{J}_m$ as a function of reservoir temperature $T/\gamma$ for different  detuning $\Delta$, driving field strength $\Omega_d$, and coupling strength $g_{mc}$. (\textbf{a}) Magnon heat current $\mathcal{J}_m$ as a function of $T_m\!=\!T$ for different driving  field strength; $\Omega_d\!=\!0$ (green dotted line), $\Omega_d\!=\!0.05\gamma$ (red dotted line), and $\Omega_d\!=\!0.1\gamma$ (blue solid line) with $\Delta\!=\!0$, $T_c\!=\!0.1$ and $g_{mc}\!=\!0.05\gamma$.  
    (\textbf{b})  Magnon heat current $\mathcal{J}_m$ as a function of $T_m\!=\!T$ for different detuning;
    $\Omega_d\!=\!0$, with $\Delta\!=\!0.05$ (green dotted line), and $\Omega_d=0.05\gamma$, $\Delta =0.05 \gamma$ (red dotted line), $\Omega_d=0.05\gamma$, $\Delta =0.5\gamma$(blue dotted line), and $\Omega_d=0.05\gamma$, $\Delta = \gamma$(black solid line). other parameters are same as (\textbf{a}). (\textbf{c}) $\mathcal{J}_m$ as a function of temperature $T/\gamma$ with $g_{mc}\!=\!0.5\gamma$, $\Omega_d\!=\!0.05$ and $\Delta\!=\!\gamma$. For forward bias case, we assumed  $T_c\!=\!0.1 \gamma$ and $T_m\!=\!T$ while the reverse bias case, $T_c\!=\!T$ and  $T_m\!=\!0.1 \gamma$.    
    (\textbf{d}) $\mathcal{J}_m$ versus temperature $T\equiv T_m$ for different magnon-photon coupling; $g_{mc}\!=\!0.05\gamma$ (red dashed line), $g_{mc}\!=\!0.3\gamma$ (blue dotted line), and $g_{mc}\!=\! 0.5\gamma$ (black solid line) with $\Omega_d\!=\!0.05 \gamma$ and $\Delta\!=\!0$. Other parameters are chosen as $\Delta_m\!=\!\Delta_c\!=\!\Delta$, $\omega_m=\omega_c=1$; and $\gamma_c\!=\!\gamma_m\!=\!\gamma\!=\!1$. 
 }
\label{fig:2}
\end{figure*}

At the steady state, the heat current flowing into the magnon or photon mode from its coupled heat reservoir can be defined as
\cite{kargi2019quantum, upadhyay2021heat, de2018reconciliation}
\begin{equation}\label{heatcurrent}
    \mathcal{J}_i \!=\!\gamma_{i} \omega_{i} (n_i -\langle \hat{i}^{\dagger} \hat{i}\rangle),
\end{equation} 
where  $\mathcal{J}_i$, $i\!=\!{m,c}$ is the rate of heat current flowing from each reservoir to the system, and $\langle \hat{i}^{\dagger} \hat{i}\rangle$ is the second moments of the operator/mode (see, Eqs. (\ref{c2moment}) - (\ref{m2moment})).
Considering symmetric reservoir couplings $\gamma_m\!=\!\gamma_c\!=\!\gamma$ as well as frequency detunings, $\Delta_m\!=\!\Delta_c\!=\!\Delta$, the heat currents are respectively calculated in a compact form as;
\begin{equation}\label{heatcurrentmag}
\mathcal{J}_{m} = -\gamma\omega_m\left(\frac{\Omega_{d}^{2}\left(\Delta^2+\frac{\gamma^{2}}{4}\right)}{\Xi}-\frac{2g_{mc}^{2}}{\gamma^{2}+4g_{mc}^{2}}\left(n_m-n_c\right)\right)
\end{equation} 
and 
\begin{equation}\label{heatcurrentphoton}
    \mathcal{J}_{c} =  -\gamma\omega_c\left(\frac{\Omega_{d}^{2}g_{mc}^{2}}{\Xi}+\frac{2g_{mc}^{2}}{\gamma^{2}+4g_{mc}^{2}}\left(n_{m}-n_{c}\right)\right).
\end{equation}
Equations \eqref{heatcurrentmag} and \eqref{heatcurrentphoton} are the exact steady-state heat currents of the driven dissipative hybrid photon-magnon when the system-reservoir couplings and the frequency detunings are symmetric. It clearly shows that there is still energy current flow for vanishing coupling between the modes due to the driving of the magnon mode. Hence, the non-conservation of heat currents, that is, $\mathcal{J}_m\!\neq\!-\mathcal{J}_c$, results from the asymmetry created by the finite magnon drive $\Omega_d\!\neq0\!$. When $\Omega_d\!=\!0$, the current is still finite due to the temperature bias, and we recover the result in \cite{landi2018quantum}. In this regime, as expected from fundamental thermodynamics that heat will always flow from hot to cold bath, the heat current $\mathcal{J}_m$ increases with temperature gradient (i.e., $n_m-n_c$) and its magnitude depends on whether the magnon bath is hotter than the photon bath or vice versa. As depicted in Fig. (\ref{figmp}), forward bias (reverse bias), the heat current $\mathcal{J}_m^f$ ($\mathcal{J}_m^r$) flows from the magnon to the photon bath (photon to magnon) when $T_m>T_c$ ($T_c > T_m$) with $T_m$ ($T_c$) denoting the temperature of the magnon (photon) bath.
The positive value of the heat current represents the flow of heat from the bath to the system and vice versa. 


We further analyse the heat current, with greater emphasis on its tunability with the driving amplitude $\Omega_d$, the detuning $\Delta$, and the temperatures of the heat reservoir. Without loss of generality, we focus on the forward-bias magnon heat current in our analysis. That is, $\mathcal{J}_m\equiv\mathcal{J}_m^f$, unless otherwise stated.
In Fig.~\ref{fig:2}(a), we show the plot of the forward bias magnon heat current against the temperature of the magnon heat bath in the absence of the detuning, $\Delta\!=\!0$, for different values of magnon driving, $\Omega_d$. The heat current increases linearly with increasing temperature and decreases with $\Omega_d$. However, when the magnon mode is driven, $\Omega_d\neq 0$, the heat current can be reversed as shown in Fig.~\ref{fig:2}(a). This is an example of how the presence of a sufficiently strong drive can flip the sign of the current, as it follows directly from Eqs. (\ref{heatcurrentmag}) and (\ref{heatcurrentphoton}). Moreover, increasing the detuning linearly improves the magnitude of the heat current for the considered temperature range (Fig.~\ref{fig:2}(b). Fig. \ref{fig:2}(c) shows that the forward bias heat current is inversely related to the corresponding reverse bias case for $\Omega_d=0$.  We further observe that by increasing the magnon-photon hybridization strength from weak to strong for non-zero magnon driving, the heat current increases, as shown in Fig.~\ref{fig:2}(d).

\begin{figure*}[!t]
\centering
\includegraphics[width=1.8 \columnwidth]{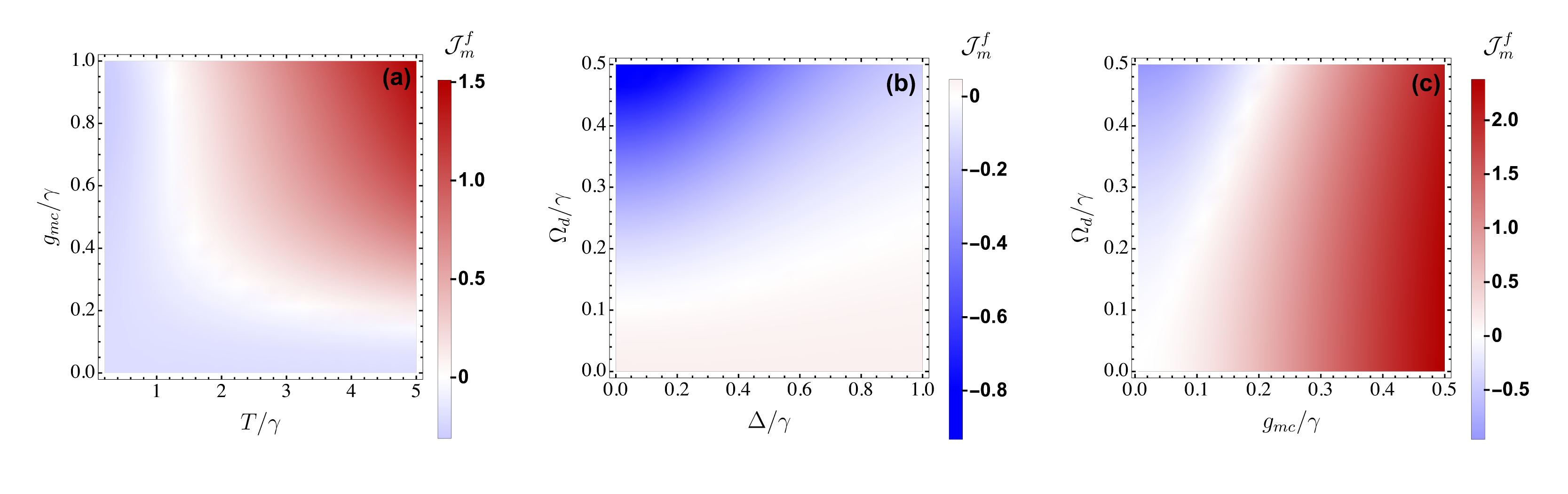} 
\caption{ \justifying Steady-state magnon heat current (forward bias) dependence on the hybrid system parameters and magnon heat reservoir temperature. (\textbf{a})  $\mathcal{J}_m$ as a function of  the coupling strength $g_{mc}/\gamma$ and magnon heat bath temperature $T_m=T$ with $\Delta\!=\!\gamma$ and $\Omega_d\!=\!0.5\gamma$. (\textbf{b}) $\mathcal{J}_m$ as a function of  driving amplitude $\Omega_d/\gamma$ and detuning $\Delta/\gamma$ with $g_{mc}\!=\!0.05\!\gamma$ and $T_m\!=\!10 \gamma$.  (\textbf{c}) $\mathcal{J}_m$ as a function of $\Omega_d/\gamma$  and magnon-photon coupling strength $g_{mc}/\gamma$ with $\Delta\!=\!0.1\gamma$ and $T_m\!=\!10 \gamma$.
Other parameters are the same as in Fig.\ref{fig:2}.}
\label{fig:4}
\end{figure*}
In summary, Fig.~\ref{fig:2} shows that (i) The magnitude of the heat current $\mathcal{J}_{m}$ depends on the parameters of the hybrid system $(\Delta\ \text{and}~  g_{mc})$, but more strongly on the driving strength of the microwave field which may alter its direction or magnitude. In the absence of magnon mode driving, the direction of heat current only changes with the interchange of heat bath temperatures (this can be confirmed from the analytical expressions in Eqs. \eqref{heatcurrentmag} and \eqref{heatcurrentphoton}).  (ii) For forward bias, in the absence of driving $\Omega_d\!=\!0$, the steady-state heat current $\mathcal{J}_{m}^f$ is positive, that is, the heat current flows from the magnon mode to the photon mode irrespective of the parameters of the system (although the presence of the drive alters the magnitude and captures regimes below zero). In the reverse bias case, the heat current $\mathcal{J}_m^r$ is negative and in the opposite direction. (iii) For very weak magnon-photon coupling $g\!\simeq\!0.005\gamma$, the negative magnitude of the heat current $\mathcal{J}_{m}$ begins to dominate. 
(iv) In the presence of the magnon mode driving, the heat current is still sustained even when the mode hybridization strength tends to vanish, $g_{mc}\!\simeq\!0$. 
Fig.~\ref{fig:4}(a) shows that high-temperature bias and strong magnon-photon hybridization are associated with enhanced heat current. In Fig. \ref{fig:4}(b), we see that the relatively strong drive can induce the heat current. The detunings for which this is most effective are functions of the system parameters, in particular the magnon-photon coupling strength, as seen in Fig.~\ref{fig:4}(b). Fig.~\ref{fig:4}(c) shows that the magnon heat current increases with $g_{mc}/\gamma$ (range of coupling strengths) and can be further tuned by variations in driving, $\Omega_d$ within the parameter regime considered. We note that the heat current is especially sensitive to the drive amplitude for moderate coupling strengths $0.1\gamma<g_{mc}<0.3\gamma$, where appropriate selection of the drive amplitude allows one to tune the amplitude and direction of the heat current. 

\subsection{Heat Rectification}
Now, we present how to utilize the tunability of the driven hybrid magnon-photon quantum system to construct a heat rectifier. The hybrid quantum system operates as a heat rectifier when there is asymmetric flow of heat current due to the dependence of its magnitude on the temperature gradient \cite{riera2019dynamically}.  Thus, the heat rectification coefficient can be  described as the asymmetry of the heat current subject to the forward and reverse directions, and it can be quantitatively described as;
\cite{joulain2016quantum, ruokola2009thermal},
\begin{equation}\label{rectification}
\mathcal{R} = \frac{|\mathcal{J}_{m}^f + \mathcal{J}_{m}^r|}{\text{max}\left(|\mathcal{J}_m^f|, |\mathcal{J}_m^r|\right)},
\end{equation}
where $\mathcal{J}_m^f$($\mathcal{J}_m^r$) correspond to the heat current in the forward (reverse) bias configuration, see Fig. \ref{figmp}. From Eq. \eqref{rectification},  $0 \leq \mathcal{R}(\mathcal{J}_{m}^f,\mathcal{J}_{m}^r) \leq 2$, the lower bound is when  there is symmetrical flow of the heat current, $\mathcal{J}_m^f\!=\!-\mathcal{J}_m^r$; while the upper bound saturates for heat fluxes that are independent of the temperature gradient, $\mathcal{J}_m^f\!=\!\mathcal{J}_m^r$. 
\begin{figure*}[!htbp]
\centering
\includegraphics[width=1.98 \columnwidth]{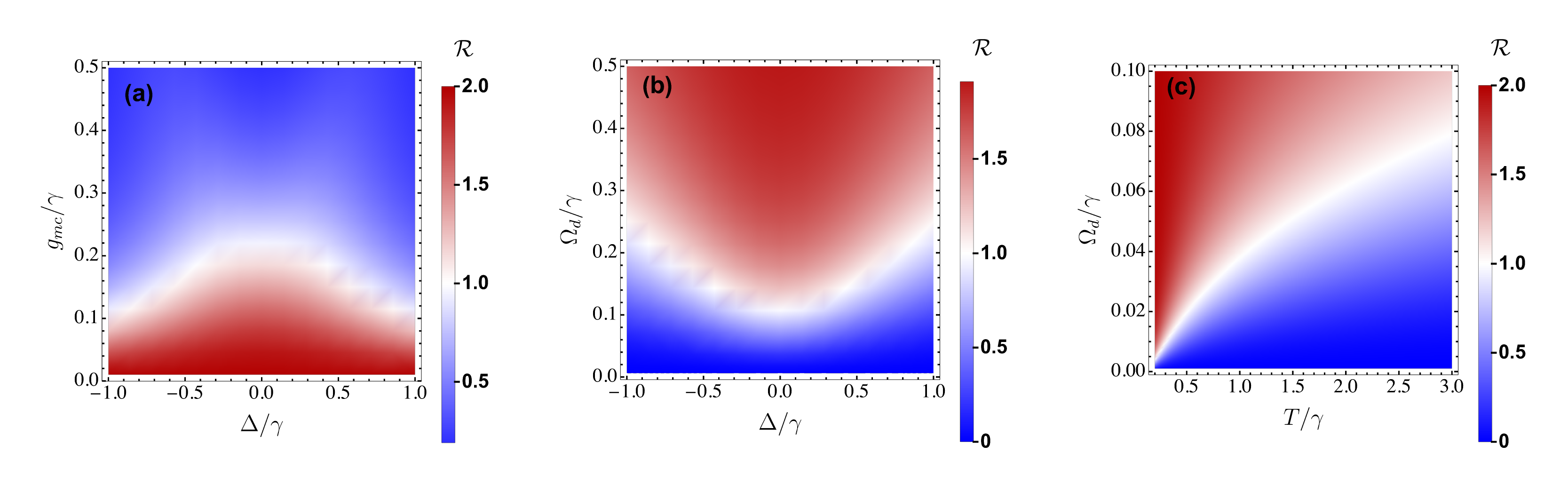} 
\caption{ \justifying (\textbf{a}) Rectification coefficient $\mathcal{R}$ as a function of magnon-photon coupling $g_{mc}$ and detuning $\Delta$ with driving amplitude $\Omega_d\!=\!0.5 \gamma$.  (\textbf{b}) Rectification coefficient $\mathcal{R}$ as a function of the strength of the magnon driving $\Omega_d$ and detuning $\Delta$ with $g_{mc}\!=\!0.05 \gamma$. (\textbf{c})   $\mathcal{R}$ as a function of the strength of the magnon driving $\Omega_d$ and magnon heat bath temperature $T_m\!=\!T$ with $g_{mc}\!=\!0.05 \gamma$,  $\Delta\!=\!0.5 \gamma$,  and $T_c\!=\!0.1 \gamma $. In (\textbf{a}) and (\textbf{b}), the temperatures are $T_c\!=\!0.1 \gamma $ (low) and $T_m\!=\!10 \gamma$ (high)}.  Other parameters are the same as in Figs. (\ref{fig:2}). 
    \label{fig:5}
\end{figure*} 
In Fig.~\ref{fig:5}, we show the dependence of the rectification coefficient $\mathcal{R}$ on the magnon-photon coupling strength $g_{mc}$, detuning $\Delta$, and magnon heat bath temperature $T_m\!=\!T$. In Fig. \ref{fig:5}(a), we see that heat rectification reduces with increasing coupling strength $g_{mc}$, in agreement with the results of the two-coupled qubit setup \cite{upadhyay2021heat}. On the other hand, at small coupling, the heat rectification is high and non-zero throughout the regime of detuning $\Delta$ considered. Figures \ref{fig:5} (b) and (c) show that when other parameters of the hybrid system and the temperature of the heat baths are fixed, adjustment of the drive amplitude $\Omega_d$ allows one to control the rectification parameter in the full range between 0 and 2. We observe the capability of the setup to attain $\mathcal{R}\!=\!1$, which indicates that the heat flux $\mathcal{J}_{m}$ is completely suppressed in one of the configurations. 
The regions with non-zero rectification correspond to regions with non-zero driving $\Omega_d$ and the maximum values of the heat rectification achieved in the regime $\Omega_d\!>\!0.5 \gamma$. In the low-temperature bias regime, the rectification factor $\mathcal{R}$ is seen to be finite, from weak to strong magnon driving amplitude, as shown in Fig. \ref{fig:5}(c). 

To gain more insight, an analytical expression for the rectification is obtained using Eq. \eqref{rectification}, and reads:
\begin{equation}\label{recanalytical}
\mathcal{R} = \Bigg |  \frac{4 \Omega_d^2 \left(\gamma^2+4 \Delta^2\right) \left(\gamma ^2+4 g_{mc}^2\right)}{2 \Omega_d^2 \left(\gamma^2+4 \Delta^2\right) \left(\gamma^2+4 g_{mc}^2\right)+g_{mc}^2 (n_c-n_m)\tilde{\mathcal{G}}_0 } \Bigg | 
\end{equation}

where $\tilde{\mathcal{G}}_0 = \left(\gamma^2+4 (g_{mc}-\Delta )^2\right) \left(\gamma ^2+4 (\Delta +g_{mc})^2 \right)$. Equation \eqref{recanalytical} shows that rectification can only be realized at non-zero drive amplitude and, it scales linearly with its intensity.
From Eq. (\ref{recanalytical}), We see that even for a sufficiently small coupling strength $g_{mc}$,  the rectification is still sustained. Moreover, in the regime of weak $g_{mc}$, we can recover the upper of bound $\mathcal{R}\!\approx\!2$.
\begin{figure}[!htbp]
\centering
\includegraphics[width=1 \columnwidth]{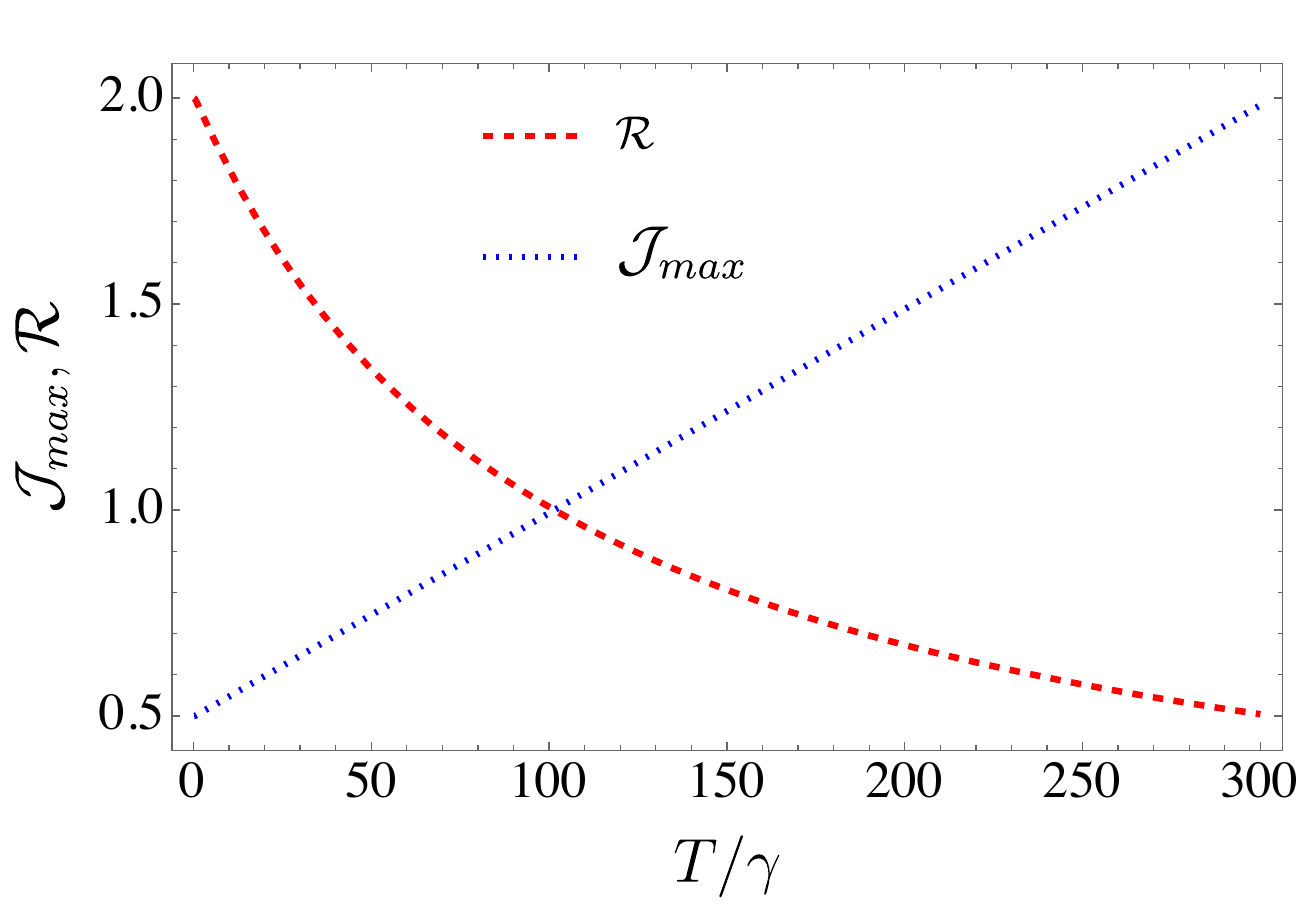} 
\caption{ \justifying Comparison between the  Rectification $\mathcal{R}$ (red curve) and $\text{max}\{ \mathcal{J}_m^f,\mathcal{J}_m^r\}/ \gamma  $ against magnon bath temperature $T_m\!\equiv\!T$. The parameters used are: $g_{mc}=0.05 \gamma$, $\Omega_d =0.5 \gamma$, $\Delta=0.5\gamma$, $T_c = 0.1 \gamma$  and $\gamma\!=\!1.$ }     
\label{figtrade}
\end{figure}
Furthermore, studies have shown that there is a trade-off between heat current and heat rectification \cite{khandelwal2023characterizing,Ordonez-Miranda2017PRE}. 
To this end, we present a comparison between the rectification and maximum heat current as a function of the magnon temperature of the heat reservoir $T_m\!\equiv\!T$ in Fig. (\ref{figtrade}). From this plot, it is explicitly demonstrated that the highest $\mathcal{R}$ occurs in the region where $\mathcal{J}_{\text{max}}\!=\!\text{max} \{\mathcal{J}_m^f, \mathcal{J}_m^r\} $ is the lowest. 
We  observe that the rectification factor (heat current) of the hybrid quantum setup is enhanced (reduced) in the low temperatures region.
This implies that a high heat current corresponds to lower rectification factors and vice versa, in agreement with the previous study on the quantum thermal device \cite{malavazi2024detuning, khandelwal2023characterizing}.

\section{Experimental Realization}\label{expt} 

Heat rectification has been experimentally demonstrated in several platforms, including phonons in carbon nanotubes \citet{chang2006solid}, and electrons in quantum dots \citet{scheibner2008quantum}, mesoscopic tunnel junctions \cite{martinez2015rectification} and suspended graphene \cite{wang2017experimental}. In \citet{senior2020heat}, rectification was achieved in a structure that integrated a superconducting quantum bit coupled to two superconducting resonators, making the two resonators of unequal length: the two resonators had in this case frequencies of 3 GHz and 7 GHz. The proposed hybrid magnon-photon system combines magnons and microwave photons in a cavity resonator, sustaining strong coupling and frequency tunability suitable for on-chip integration and coherent control of energy exchange. The driven magnon-photon system offers distinct advantages over other approaches by enabling active control of heat rectification through external driving fields, which enhances tunability. This modulation supports asymmetric steady-state heat currents, making it highly useful for quantum thermal management. Additionally, the resonance frequency of the magnon mode is tunable by a magnetic field, which could serve as an additional resource to achieve heat rectification (asymmetry) in the setup. Thus, compared to existing platforms, driven magnon-photon systems combine active control through optical, thermal, and magnetic knobs, scalability, and strong hybrid coupling to realize quantum heat rectifiers.

A possible implementation setup for a magnon-photon system as a quantum heat rectifier would include a three-dimensional (3D) microwave cavity and a well-polished Yttrium iron garnet (YIG) sphere; the YIG functions as the magnon cavity. The cavity is a box made of highly conductive copper to achieve a high-quality factor at room temperature \cite{Zhang2014PRL}. Then, the YIG sphere is placed inside the microwave cavity and biased with a static magnetic field. The magnetic components of the microwave field are perpendicular to the bias field which induces the spin-flip and thus excites the magnon mode in YIG. The lowest order ferromagnetic resonance mode consists of a uniform collective mode, in which all the spins precess in phase. This mode has the highest coupling strength, since the microwave magnetic field around the YIG sphere is approximately uniform as the wavelength $\lambda_{mw}\!\gg\!R$, where $R$ is the radius of the YIG sphere. The frequency of the uniform magnon mode linearly depends on the bias field: $\omega_m\!=\!\gamma_{gr} |B_0| + \omega_{m,0}$, where $\gamma_{gr}$ is the gyromagnetic ratio and $\omega_{m,0}$ is determined by the anisotropy field. The bias magnetic field is tunable in the range of $0 \text{--} 2$ Tesla, corresponding to a magnon frequency from a few hundred MHz to about $50\,\text{GHz}$ \cite{Zhang2014PRL}. To achieve the photon-magnon interaction, the bias field is adjusted such that the magnon is near resonance with the photon mode. The strongest interaction coupling strength is obtained by placing the YIG sphere in the position with the maximum microwave magnetic field \cite{crescini2020magnon, Sun2021PRL, Zhang2014PRL}, which has been implemented in recent experiments \cite{Tabuchi2014PRL, Zhang2014PRL}.  
This demonstrates the fact that the setup is a useful resource for the implementation of controllable quantum heat rectifiers. Finally, experimentally feasible parameters to implement the scheme are as follows: the strong coupling between the microwave photons and the magnon, with the coupling strength $g_{mc}/2\pi = 10.8 \text{MHz}$, the dissipation rates of the microwave photon and the magnon are respectively, $\gamma_c/2\pi\!=\!2.67 \text{MHz}$ and $\gamma_m/2\pi\!=\!2.13 \text{MHz}$, the resonance frequency of the photon $\omega_c/2\pi\!=\!10 \text{GHz}$ \cite{Zhang2014PRL, li2018magnon}, and the magnon drive is given as follows; $\Omega_d=\frac{\sqrt{5}}{2}\gamma_{gr}\sqrt{N} B_{0}$, where $\gamma_{gr} = 28 \text{GHz/T}$ is the gyromagnetic ratio, $N= \rho V$ is total number of spins and $\rho=4.22 \times 10^27 \text{m}^{-3}$ is the spin density of the YIG , and $V$ is the the volume of the sphere \cite{li2018magnon}. Furthermore, we add that the results presented here are only valid in the weak-coupling regime. 

\section{CONCLUSIONS} \label{conclusion}
We have investigated heat transport and heat rectification in a driven dissipative magnon-photon system. The system is in contact with two fixed temperature reservoirs. Employing the Lindblad master equation formalism for an open quantum system, we derived closed-form analytical expressions for the steady-state heat current and identified the key mechanisms responsible for asymmetry in heat transport. The external magnon drive was found to play a crucial role not only in inducing asymmetry but also in serving as a powerful tuning parameter for the thermal response of the hybrid quantum system: In the weak magnon-photon coupling regime, the drive was demonstrated to enable a reversal of the heat current direction and the tunability of the heat rectification in its full physical range. 
Our results show that a finite heat current and rectification persist even in the vanishing-coupling limit, entirely due to the considered drive. This highlights a new operational regime for heat rectification.

These results highlight the significant potential of driven hybrid magnon-photon systems in the advancement of the design of quantum thermal devices, including thermal rectifiers, thermal diodes, and thermal transistors. Furthermore, our work opens new avenues for exploring quantum control strategies in heat transport, offering a promising platform for future research into quantum thermal machines and energy-efficient quantum technologies.

\section*{Acknowledgements}
COE and NA acknowledge support from the Long Term Research Grant Scheme (LRGS) with grant number LRGS/1/2020/UM/01/5/2 (9012-00009), provided by the Ministry of Higher Education of Malaysia (MOHE). KS acknowledges the National Science Centre, Poland, project number 2020/39/I/ST3/00526. Part of this project was carried out during COE's visit to  Nicolaus Copernicus University (NCU) in Torun, and he acknowledges the support of the NCU's Research University Centre of Excellence "From Fundamental Optics to Applied Biophotonics". 

\bibliographystyle{apsrev4-2}
\bibliography{References}
\end{document}